\documentclass[aps,superscriptaddress,twocolumn,groupedaddress]{revtex4}

\ifx\pdfoutput\undefined\usepackage{graphicx}\else\usepackage{graphicx}\fi
\usepackage{color}%
\usepackage[final]{LLNdef}

\begin{document}



\title{Nanometers-thick self-organized Fe stripes: bridging the gap between surfaces and magnetic materials}


\author{O. Fruchart}
\email[]{Olivier.Fruchart@grenoble.cnrs.fr}
\author{M. Eleoui}%
\author{J. Vogel}%
\affiliation{Laboratoire Louis N\'{e}el (CNRS, UPR5051), BP166, F-38042 Grenoble Cedex 9, France}%
\author{P. O. Jubert}%
\affiliation{IBM Research, Z{\"u}rich Research Laboratory - S{\"a}umerstrasse 4, 8803 R{\"u}schlikon, Switzerland}%
\author{A. Locatelli}%
\author{A. Ballestrazzi}%
\affiliation{ELETTRA S.C.p.A., I-34012 Basovizza, Trieste, Italy}%


\date{\today}

\def\RT{RT}%
\def\SO{SO}%

\def\few{Fe/W}%
\def\fewmo{Fe/W/Mo}%
\def\Ip{I^+}%
\def\Im{I^-}%

\begin{abstract}
We have fabricated $\thicknm5$-high Fe(110) stripes by self-organized~(\SO) growth on a slightly
vicinal R(110)/\saphir$(11\overline20)$ surface, with R=Mo,~W. Remanence, coercivity and domain
patterns were observed at room temperature~(\RT). This contrasts with conventional \SO\ epitaxial
systems, that are superparamagnetic or even non-magnetic at~\RT\ due to their flatness. Our
process should help to overcome superparamagnetism without compromise on the lateral size if \SO\
systems are ever to be used in applications.
\end{abstract}


\maketitle


Arrays of epitaxial nanometer-sized~(\thicknm{1-50}) magnetic structures can be grown by
self-organization~(\SO). However such structures are superparamagnetic or even non-magnetic at
room temperature~(RT)\cite{bib-PAD99b,bib-HAU98b,bib-GAM02}. Indeed the energy barrier opposing
spontaneous magnetization flipping roughly scales with $KV$, with $K$ the magnetic anisotropy per
unit volume, and $V$ the system's volume. 3D clusters of similar lateral size can overcome
superparamagnetism at \RT\ by increasing $K$\cite{bib-SUN00}. This seems not sufficient in
epitaxial \SO\cite{bib-GAM02,bib-OHR01,bib-GAM03} because \SO\ deposits are generally very flat,
implying a very small~$V$. Therefore, beating superparamagnetism in \SO\ deposits without
compromising on the lateral density seems to imply increasing their thickness~$t$.

One way to force \SO\ deposits to grow vertically and overcome superparamagnetism at \RT\ is
sequential deposition\cite{bib-FRU99d,bib-FRU02b}. We proposed a second route, that consists in
annealing a thin continuous film deposited on a vicinal surface to form an array of several
atomic layers~(\AL)-thick stripes\cite{bib-FRU02b,bib-FRU02c}. In the early reports, concerning
Fe/Mo(110) stripes, a stable thickness $t=\thickAL{6}$~($\sim\thicknm{1.2}$) was observed above
\unit[1-2]{\AL s} of wetting. Yet this was not thick enough to observe static coercivity at \RT,
which could be obtained only for multidisperse assemblies of islands and stripes, thicker on the
average. In this Letter we report the growth of thicker stripes, in the case of Fe/W(110):
$t\sim\thickAL{25}$~($\sim\thicknm5$). Such stripes display at \RT\ functional features of
magnetic materials: coercivity, remanence and domains, unlike conventional \SO\ systems. The
microscopic origin of the self-organization process is also unravelled.

\begin{figure}[b]
  \includegraphics[width=8.5cm]{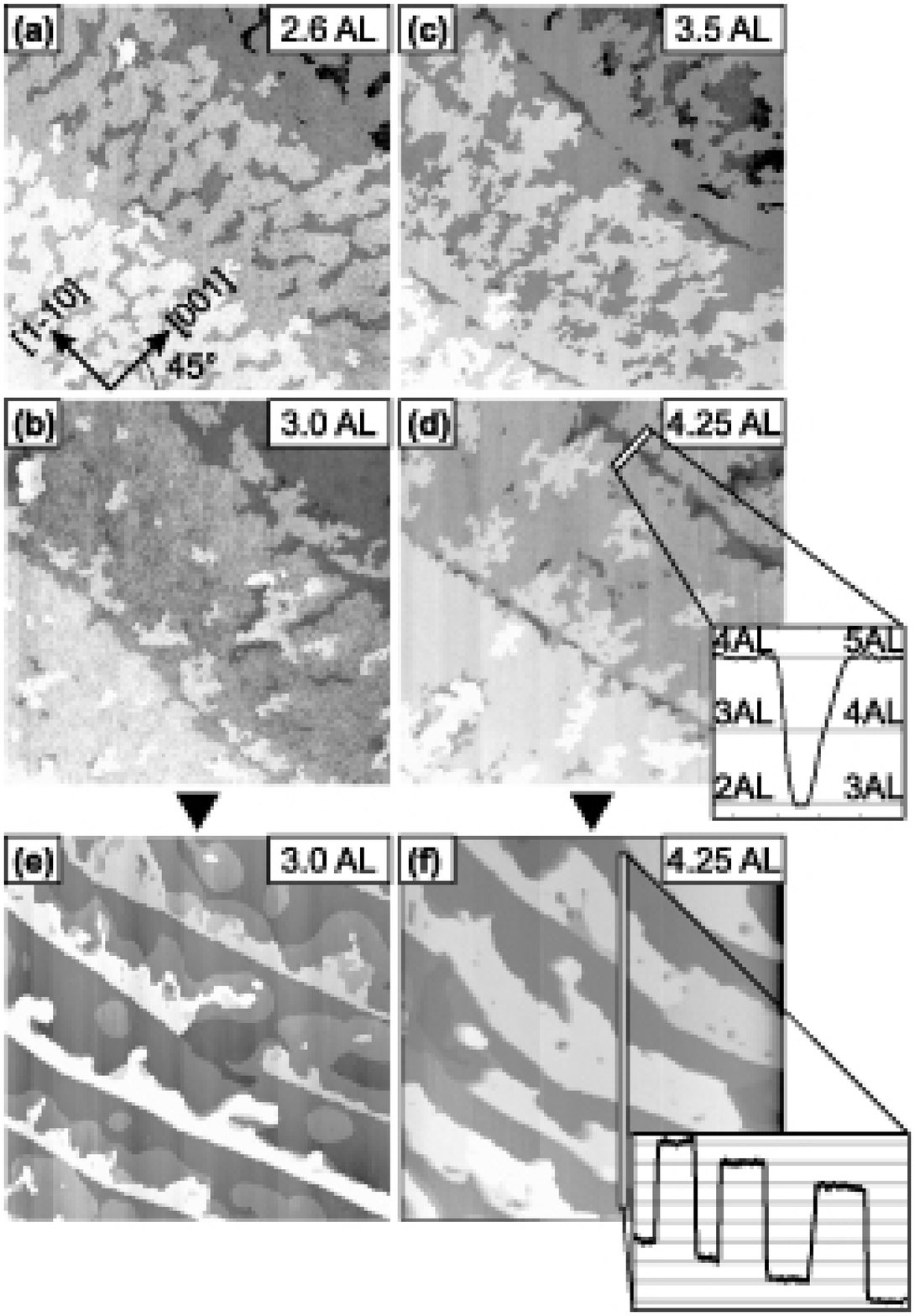}%
  \caption{\label{fig-growth}\dataref{POJ, STM PO105}STM pictures of Fe/Mo(110)
    after growth at $\unit[150]{^\circ\mathrm{C}}$~(a-d, $\unit[400\times400]{nm}$) and after
    annealing at $\unit[450]{^\circ\mathrm{C}}$~(e-f, $\unit[800\times800]{nm}$), for varying Fe
    nominal thickness~(see upper-right labels).
    The insets show cross-sections with integer \AL s sketched by gray lines.}
\end{figure}

The samples are epitaxially grown by pulsed laser deposition in a multi-chamber ultra-high vacuum
setup~(base pressure $\unit[\scientific7{-9}]{Pa}$), with \insitu\ STM, RHEED and Auger
spectroscopy\cite{bib-FRU03d}. Commercial $(11\overline{2}0)$ sapphire wafers with a residual
miscut angle $\epsilon<\unit[0.1]{\deg}$ are buffered with refractory metal films~(Mo or W,
$\lesssim\thicknm{10}$-thick), whose surface consists of an array of atomically-flat terraces of
width $\sim\thicknm{200}$, separated by mono-atomic steps\cite{bib-FRU02c}. Fe is then deposited
at $\tempdegC{150}$ and annealed at $\tempdegC{400-450}$, covered with \thicknm{1} Mo for
controlling the magnetic anisotropy, and finally capped by $\thicknm4$~Al as a protection against
oxidation. AFM~(PSI Autoprobe CP) and hysteresis loops~(QD MPMS-XL) were performed \exsitu.
Samples were then dc-demagnetized \exsitu\ with the field applied perpendicular to the stripes.
Magnetic and chemical imaging was performed under zero external field using the Spectroscopic
Photo Emission- and Low Energy Electron Microscope (SPELEEM), operational at the Nanospectroscopy
beamline, at the Elettra synchrotron radiation facility in Trieste, Italy\cite{bib-LOC03}.
Element-selective magnetic contrast was obtained by combining energy filtered PEEM with X-ray
Magnetic circular dichroism~(XMCD). The circularly-polarized X-ray beam was shone on the sample
at an angle of incidence of \unit[16]{\deg} and parallel to the in-plane $[1\overline10]$ of the
sample, \ie\ roughly perpendicular to the stripes. The photon energy was tuned to the
Fe~$\mathrm{L}_3$ edge. The magnetic contrast $(\Ip-\Im) / (\Ip+\Im)$, or XMCD asymmetry, is
proportional to the projection of the magnetization along the X-ray beam direction, where for
each pixel $\Ip$ and $\Im$ are the intensity acquired with opposite helicity of the photon beam.
The chemical contrast is given by $(\Ip+\Im)/2$.

In conventional step decoration processes, the stripes have a height comparable to that of
mono-atomic steps\cite{bib-SHE97d,bib-HAU98b,bib-GAM02}. \figref{fig-growth} reveals the
microscopic mechanism allowing the array of mono-atomic steps to serve as a template for the
self-organization of stripes much thicker than the steps themselves. The frontier between the
low-temperature growth mode with increasing kinetic roughness\cite{bib-ALB93} and the
high-temperature Stranski-Krastanov growth mode with dots formation\cite{bib-FRU01b} is
$\tempdegC{150}$. At this temperature the growth of Fe proceeds layer-by-layer on terraces, but
grooves with a depth increasing with nominal Fe thickness are observed in register with the
initial array of atomic steps\bracketsubfigref{fig-growth}{b}, probably resulting from stress
effects\cite{bib-MUL01} and the connection of interface dislocation arrays across
steps\cite{bib-MUR02}. Annealing Fe/W(110) films is known to yield stripes aligned along
$[001]$\cite{bib-SAN98b}. Here the grooves drive nucleation, yielding a self-organized array of
stripes roughly aligned along $[1\overline10]$ for this wafer~(see FIG.\ref{fig-growth}a). The
straighter side of the stripes is the one lying along the steps of the buffer
layer~(\figref{fig-growth}{e-f}). As grooves along buried steps have been observed in other
systems\cite{bib-CHE01}, the self-organization process described here is expected to be of
broader validity than solely for Fe/bcc(110).

\begin{figure}[h]
  \includegraphics[width=8.5cm]{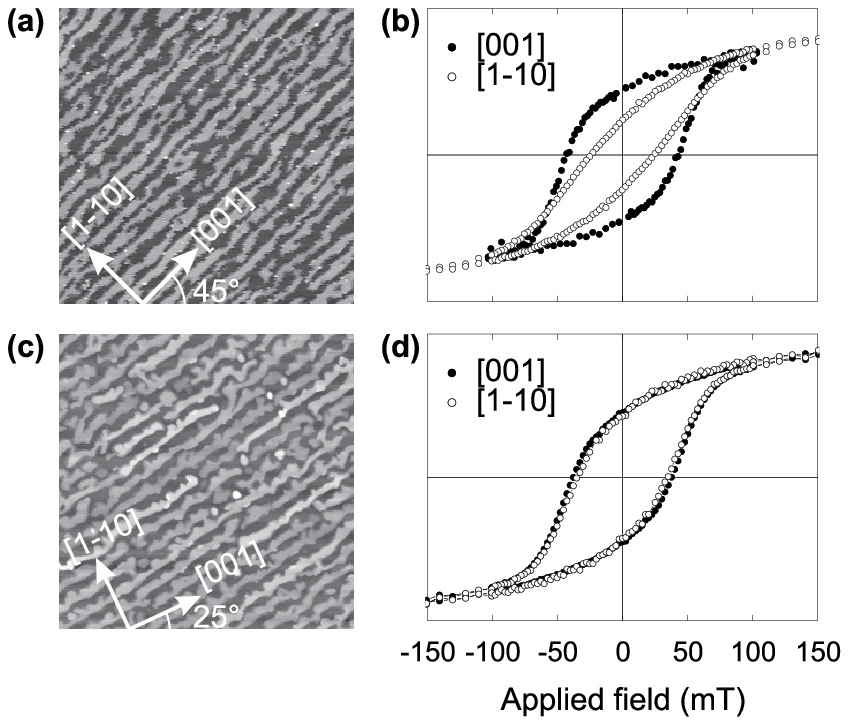}%
  \caption{$\unit[5\times5]{\mu\mathrm{m}}$ AFM pictures of samples
  (a)~\few\dataref{Fru101, AFM240039}\ and (c)~\fewmo\dataref{Fru102b, AFM29004b}~(see text for definition).
  The in-plane lattice directions are shown. In-plane hysteresis loops of samples
   (b)~\few\ and (d)~\fewmo. \label{fig-afm-magn}}
\end{figure}

In the case of Fe/W(110)/\saphir$(11\overline20)$~(\subfigref{fig-afm-magn}a, sample called \few\
in the following) the stripes display a sharp distribution function of height, centered around
$t\sim\thicknm{5.5}$ \stress{independent of nominal Fe thickness}, a value that is much higher
than for Fe/Mo(110)~($\thicknm{1.2}$, see above). For a second sample a $\thicknm{1}$-thick
pseudomorphic W film was deposited onto a Mo(110) buffer layer; the absence of Mo segregation
towards the surface was checked with a quantitative Auger analysis. For this composite buffer
layer, Fe stripes with a monodisperse thickness $t\sim\thicknm{4.3}$, again independent of
nominal Fe thickness, are observed~(\subfigref{fig-afm-magn}c, sample called \fewmo). This
suggests that both the lattice parameter of the buffer layer and the interfacial energy influence
the stripe thickness. The origin of this metastable thickness is under investigation. For both
samples the miscut angle is $\unit[0.04]{\deg}$, yielding a terrace width of $\thicknm{290}$.

\subfigref{fig-afm-magn}b shows \RT\ hysteresis loops of sample \few\ along two in-plane
directions. $[1\overline10]$ is a hard magnetic axis and $[001]$ is an easy axis, with
significant remanence and a coercivity of $\unit[43]{mT}$. The coercivity at $\tempK{10}$ is only
slightly increased to $\unit[49]{mT}$, confirming the weak effect of temperature that results
from the large activation volume expected in thick stripes. This contrasts with the
superparamagnetic behavior of conventional \SO\ systems. \subfigref{fig-afm-magn}d shows \RT\
hysteresis loops of sample \fewmo. This time both in-plane axes are magnetically similar, with
again significant remanence and a mean coercive field of $\unit[36]{mT}$~(we do not discuss the
origin of anisotropy in these samples, which is a complex balance between bulk, magneto-elastic,
interface\cite{bib-GRA93,bib-OSG95b,bib-FRU99c} and shape anisotropies).

\begin{figure}[t]
  \includegraphics[width=8.5cm]{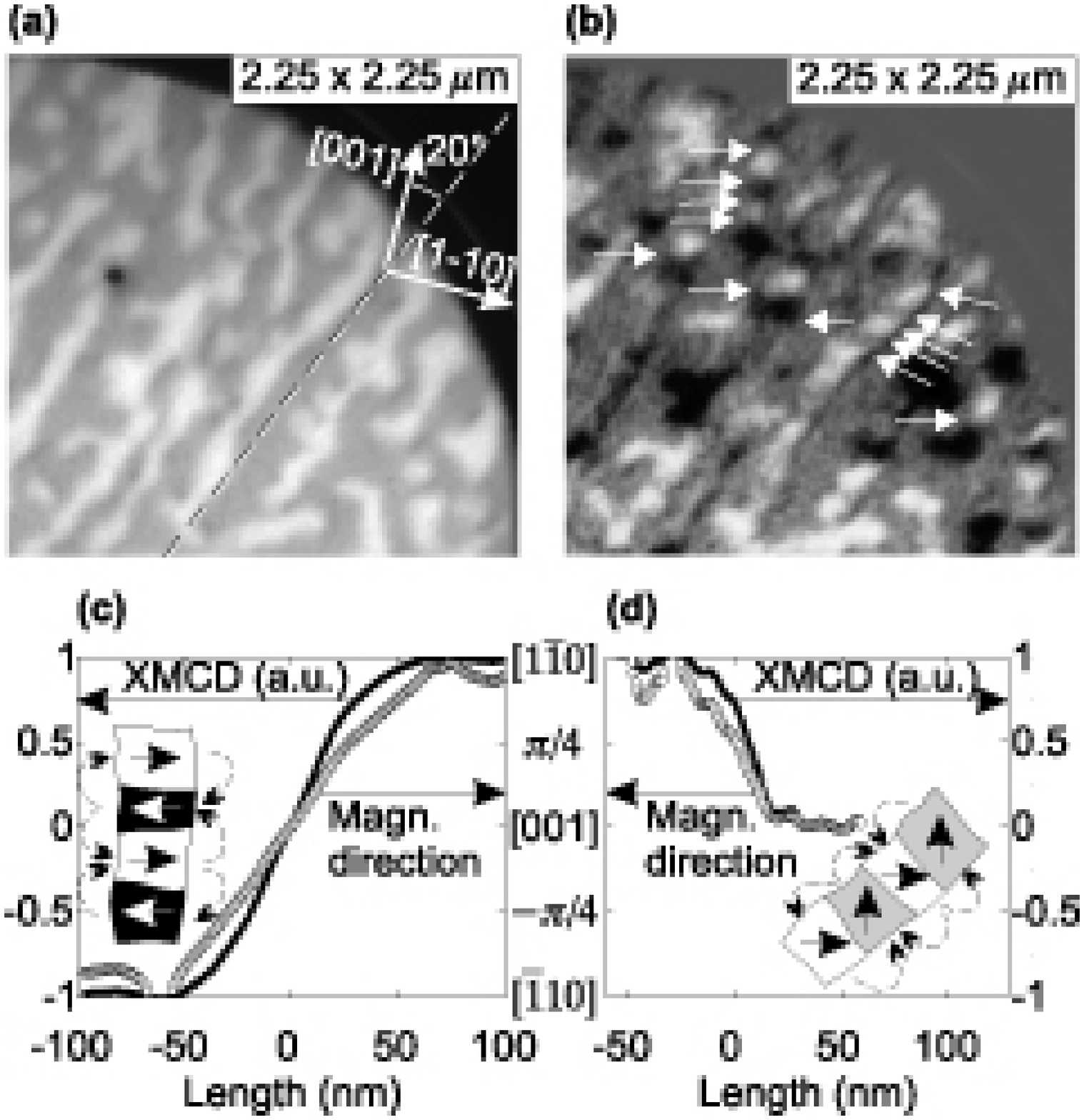}%
  \caption{\dataref{\figref{fig-peem}: 2002-12-04-023; m\^{e}me zone que 2002-12-04-053, imag\'{e}e plus tard en vue
2.5microns}(a)~chemical and (b)~magnetic $\unit[2.25\times2.25]{\mu\mathrm{m}}$ PEEM images of
  the same area of sample \fewmo. In (b) arrows pointing to the right~(left) indicate
  some $\unit[180]{^\circ}$ ($\unit[90]{^\circ}$) domain walls. (c-d)~
  profiles of $\unit[180]{^\circ}$~(c) and $\unit[90]{^\circ}$~(d) walls, obtained as the average
  of several cross-sections in (b) for (c), and in images with a \textsl{total} field of view
  of $\unit[2.5\times2.5]{\mu\mathrm{m}}$ (not shown here) for $\unit[90]{^\circ}$ walls.
  The magnetization direction is calculated as $\arccos(\mathrm{XMCD})$. Insets: sketches of the
  flux closure domains indicated by a set of arrows in~(b). The dotted-line arrows sketch the flux closure. \label{fig-peem}}
\end{figure}

\figref{fig-peem} shows PEEM images of sample~\fewmo, with a nominal thickness slightly smaller
than in \subfigref{fig-afm-magn}c, yielding narrower and more irregular stripes~(see chemical
contrast on \subfigref{fig-peem}a). In \subfigref{fig-peem}b) light (resp. dark)  areas correspond
to magnetic domains pointing along $[1\overline10]$ (resp.$[\overline110]$)
\bracketsubfigref{fig-afm-magn}c, \ie\ roughly perpendicular to the stripes. Grey areas
correspond to domains along $[001]$, \ie\ roughly parallel to the stripes, or to the non-magnetic
material between the stripes . These domain patterns are similar to those observed at the
macroscopic scale\cite{bib-HUB98b}. They arise to satisfy both stray field flux-closure and
charge-free domain walls. Depending on the local orientation of the stripes with respect to the
crystallographic directions, $\unit[180]{\deg}$ and $\unit[90]{\deg}$ domain walls are
observed\bracketsubfigref{fig-peem}{c-d}. We follow the usual definition of the width $\lambda$
of a wall as the width of the linear asymptote to the plot of magnetization angle versus length.
We found\bracketsubfigref{fig-peem}{c-d} $\lambda_{180}=\thicknm{110}$ and
$\lambda_{90}=\thicknm{50}$, respectively, satisfying $\lambda_{180}\sim2\lambda_{90}$ as
expected. Note that the numerical value of $\lambda_{90}$ is however close to the expected
lateral resolution of the microscope~(a few tens of nm in PEEM mode).

To conclude we have unravelled a microscopic mechanism that allows a vicinal surface to serve as
a template for the self-organized growth of stripes displaying a monodisperse thickness of up to
$\thicknm{5.5}$. This growth process allows one to overcome superparamagnetism without
compromising on the lateral density. Thus for the first time self-organized magnetic
nanostructures were observed to display at room temperature usual features of bulk materials:
remanence, coercivity and domain patterns.

\begin{acknowledgments}
We are grateful to Ph.~David and V.~Santonacci for their technical support, and J.~Camarero,
F.~Scheurer and Y.~Samson for preliminary magnetic measurements. This work was partly funded by
\textsl{R\'{e}gion Rh\^{o}ne-Alpes}~(Project Emergence 2001) and the French ministry of Research~(ACI
Nanostructures 2000).
\end{acknowledgments}


\bibliography{fruche3}

\end{document}